%% file: unruh-fd.tex
\documentstyle[12pt]{article}
\def\bmi#1{\mbox{\boldmath $#1$}}

\makeatletter
\@addtoreset{equation}{section}

\makeatother

\begin{document}
\sloppy
\sloppy
\sloppy

\begin{flushright}{UT-839\\March, 1999\\ 
%unruh-fd.tex
}

\end{flushright}
\vskip 0.5 truecm

\vskip 0.5 truecm

\begin{center}
{\large{\bf  Fluctuation-dissipation theorem and 
the Unruh effect of scalar and Dirac fields}}
\end{center}
\vskip .5 truecm
\centerline{\bf Hiroaki Terashima}
\vskip .4 truecm
\centerline {\it Department of Physics,University of Tokyo}
\centerline {\it Bunkyo-ku,Tokyo 113-0033,Japan}
\vskip 0.5 truecm

\begin{abstract}
We present a simple and systematic method to
calculate the Rindler noise, which is relevant
to the analysis of the Unruh effect, 
by using the fluctuation-dissipative theorem.
To do this, we calculate the dissipative coefficient
{\em explicitly} from the equations of motion
of the detector and the field.
This method gives not only the correct answer
but also a hint as to the origin of the apparent
statistics inversion effect.
Moreover, this method is generalized to the Dirac field,
by using the fermionic fluctuation-dissipation theorem.
We can thus confirm that the fermionic fluctuation-dissipation
theorem is working properly.
\end{abstract}

\section{Introduction}
\input{intro.tex}

\section{Rindler Noise of Real Scalar Field}
\input{scalar.tex}

\section{Generalization to Complex Scalar Field}
\input{complex.tex}

\section{Generalization to Dirac Field}
\input{dirac.tex}

\section{Conclusion}
\input{conclude.tex}

\section*{Acknowledgment}
The author thanks K. Fujikawa for introducing him
to this subject and helpful discussions.
The author also thanks Y. Nomura for discussions and
encouragement.

%\bibliographystyle{prsty}
%\bibliography{phys}

\input{ref.tex}
\end{document}

%% file: intro.tex
For a uniformly accelerated observer
in the flat spacetime,
the ordinary Minkowski vacuum,
which is defined by using the ``frequency''
with respect to the Minkowski time,
can be seen as a thermal bath of particles~\cite{Unruh76}.
(The temperature of the thermal bath is $\hbar a/2\pi$
where $a$ is the acceleration of the observer.)
These particles are considered as the Rindler particles
which are defined by using the ``frequency''
with respect to the Rindler time~\cite{Fullin73,Davies75,UnrWal84}.
This effect is called the Unruh effect.

In order to see this effect,
it is convenient to introduce a model of a ``particle detector''
called the DeWitt detector~\cite{DeWitt79}.
The DeWitt detector is a point-like object with internal
energy levels and moves along the world line $x(\tau)$
where $\tau$ is the proper time of the detector.
The internal energy levels of the detector are $\{E_i\}$
and corresponding eigenstates are $\{|E_i\rangle\}$.
That is,
\begin{equation}
   H_D|E_i\rangle=E_i|E_i\rangle,
\end{equation}
where $H_D$ is the Hamiltonian of the internal structure 
of the detector.
Also, the detector has an internal degree of freedom
(``monopole moment'') $Q(\tau)$.
The time evolution of $Q(\tau)$ in the Heisenberg picture is
\begin{equation}
   Q(\tau)=e^{iH_D\tau/\hbar}Q(0)\;e^{-iH_D\tau/\hbar}.
\label{Qtau}
\end{equation}
And then, the detector is linearly coupled 
to a field $\phi(x)$ via this monopole.
The interaction is described by the interaction Lagrangian
\begin{equation}
  L_{int}=Q(\tau)\phi(\tau),
\end{equation}
where $\phi(\tau)\equiv\phi(x(\tau))$ is the value of
the field along the world line of the detector.
(To be precise, we need the adiabatic switching.)

The measurement process by this detector is as follows.
Suppose that, at $\tau=-\infty$, the detector is
in the ground state $|E_0\rangle$ and
the field is in the Minkowski vacuum $|0_M\rangle$.
After the detector-field interaction is switched on,
the detector would not remain in $|E_0\rangle$,
but would make a transition to an excited state.
It means that the detector ``detects'' some particles.

The transition amplitude for the detector-field system
to be found in $|E_1,\psi\rangle$ at $\tau=\infty$ is
given by first order perturbation theory as
\begin{equation}
  i\langle E_1,\psi|\int^\infty_{-\infty} d\tau\;
   Q(\tau)\phi(\tau)|E_0,0_M\rangle.
\end{equation}
(Here, it has been assumed that the matrix element of $Q$
is sufficiently small enough for the perturbation theory
to be appropriate.)
By Eq.(\ref{Qtau}),
this can be written as
\begin{equation}
  i\langle E_1|Q(0)|E_0\rangle
   \int^\infty_{-\infty} d\tau\;e^{i(E_1-E_0)\tau/\hbar}
   \langle\psi|\phi(\tau)|0_M\rangle.
\end{equation}
Therefore, after summation over all the final 
states of the field $|\psi\rangle$, the transition rate 
(the transition probability per unit proper time) 
from $E_0$ to $E_1$ is
\begin{equation}
  |\langle E_1|Q(0)|E_0\rangle|^2\;
  {\cal F}\left(\frac{E_1-E_0}{\hbar}\right),
\end{equation}
where
\begin{equation}
    {\cal F}(\omega)=\int^\infty_{-\infty}d(\tau-\tau')
    \;e^{-i\omega(\tau-\tau')}g(\tau-\tau')
\end{equation}
and
\begin{equation}
   g(\tau-\tau')=\langle0_M|\phi(\tau)\phi(\tau')|0_M\rangle.
\end{equation}

Thus, the transition rate of the DeWitt detector is 
proportional to the ``response function'' ${\cal F}(\omega)$
which depends only on the field and the world line of the detector
but not on the internal structure of the detector.
Note that we may regard $g(\tau-\tau')$ as a kind of noise,
``the quantum noise in the Minkowski vacuum along
the world line $x(\tau)$'' and the response function as
the power spectrum of this noise.

For the Unruh effect, the relevant noise is the Rindler noise,
where $x(\tau)$ is a uniformly accelerated world line.
In two dimensions,
the power spectrum of the Rindler noise 
(the response of the detector) is exactly
those of the thermal noise in the thermal bath.
However, in other dimensions, 
it was shown by Takagi~\cite{Takagi86} that
there are some differences
between the Rindler noise and the thermal noise.
Specially, the Rindler noise exhibits the phenomenon of
the apparent inversion of statistics in odd dimensions.

In this paper, we present a simple and systematic method 
to reproduce these results by using the 
fluctuation-dissipation theorem~\cite{CalWel51,Kubo57,FeyVer63}
which is a basis of statistical mechanics for irreversible
processes when the systems are slightly away
from thermal equilibrium.
This theorem states the relation between the spontaneous
fluctuation of fields in thermal equilibrium 
and the irreversible dissipation.
Although the fluctuation-dissipation theorem has been
formulated by various authors, we adopt the formulation 
by Callen and Welton~\cite{CalWel51}, 
because their formulation is intuitively
understandable and appealing. 
They showed that a general form of fluctuation-dissipation
theorem covers a wide range of phenomena such as 
the Einstein relation for Brownian motion, 
the Nyquist formula for voltage fluctuation in conductors,
and the Planck distribution for photons. 

Of course, the ``fluctuation-dissipation point of view''
was already pointed out~\cite{Ooguri86,Takagi86}.
However, the previous discussions concentrated on 
the vacuum expectation value of the commutator
and anticommutator of the fields 
in connection with the Huygens' principle.
Although these commutators are related to the dissipation,
the notion of ``dissipation'' is not quite clear.
In addition, the calculations involved 
in obtaining these commutators are quantum and 
essentially the same as those 
to obtain the ``fluctuation'' directly.

In contrast, in our application of
the fluctuation-dissipation theorem
to the Unruh effect, we calculate the dissipative coefficient
{\em explicitly} from the equations of motion
of the detector and the field.
Then, by virtue of the theorem,
we can immediately obtain the Rindler noise.
All the calculation we need is {\em classical} and
thus our calculation is completely {\em different}
from the previous ones.
By this calculation, we can get 
not only the correct answer
but also a hint as to the origin of the apparent
statistics inversion effect.

Moreover, our method is generalized to a Dirac field,
by using the fluctuation-dissipation theorem
of a fermionic operator~\cite{FujTer98}.
In the context of condensed matter physics,
the fermionic fluctuation-dissipation theorem
may not be directly applicable, because we usually measure
bosonic quantities such as voltage or electric current.
In this paper, we show that the fermionic 
fluctuation-dissipation theorem is indeed {\em applicable}
to the Unruh effect and works {\em properly}.

To our knowledge, the present approach to the Unruh effect
by using the bosonic and fermionic fluctuation-dissipation theorem
as a corner stone has not been discussed before.

%% file: scalar.tex
\subsection{Thermal Noise}
In order to see how the fluctuation-dissipation theorem
works not only for the thermal noise but also
for the Rindler noise to be discussed later, 
we first consider the thermal noise.
The system consists of a real scalar field and a detector
in $n$ dimensional flat spacetime.
The detector is at rest and linearly coupled with
the scalar field through 
an internal degree of freedom $Q(t)$, there.
The scalar field is initially
in thermal equilibrium at temperature $T$.
Thus, the action of the total system is
\begin{equation}
S = S_0(Q)+S_{int}(Q,\phi)+S_0(\phi),
\end{equation}
where
\begin{eqnarray}
S_0(Q)  &=& \int dt\;L(Q,\dot{Q}), \\
S_{int}(Q,\phi)&=&\int dtd\vec{x}\;Q(t)\,\phi(\vec{x},t)\,
               \delta(\vec{x}-\vec{x}_0) \nonumber \\
        &=& \int dt\;Q(t)\,\phi(\vec{x}_0,t),\label{TSSint} \\
S_0(\phi) &=& \int dtd\vec{x}\; \frac{1}{2} \,
  \left[ (\partial_t\phi)^2-\sum_{i=1}^{n-1}
         (\partial_i\phi)^2-m^2\phi^2 \right],\label{TSS0}
\end{eqnarray}
and $\vec{x}_0$ is the position of the detector.
($\vec{x}$ stands for an $(n-1)$-dimensional vector.)
Here, we do not need an explicit form of
$L(Q,\dot{Q})$ which is the Lagrangian for
the (unperturbed) detector.

From this action, we can derive the equations of motion
\begin{eqnarray}
  \left(\frac{\delta S_0}{\delta Q}\right)
 +\int d\vec{x}\;\phi(\vec{x},t)
    \,\delta(\vec{x}-\vec{x}_0) &=& 0, \\
\partial_t^2\phi-\sum_i\partial_i^2\phi+m^2\phi 
    - Q(t)\,\delta(\vec{x}-\vec{x}_0)&=&0,
\end{eqnarray}
where
\begin{equation}
-\left(\frac{\delta S_0}{\delta Q}\right)=
 \frac{d}{dt}\left(\frac{\partial L}{\partial \dot{Q}}\right)
 -\left(\frac{\partial L}{\partial Q}\right).
\end{equation}

By Fourier transformations,
\begin{eqnarray}
\tilde{\phi}(\vec{k},\omega) &\equiv& 
   \int dtd\vec{x}\;\phi(\vec{x},t)\,
      e^{i\vec{k}\vec{x}-i\omega t}, \\
\tilde{Q}(\omega) &\equiv&
   \int dt\;Q(t)\,e^{-i\omega t},
\end{eqnarray}
these equations become
\begin{eqnarray}
 \left(\widetilde{\frac{\delta S_0}{\delta Q}}\right)
 +\int \frac{d\vec{k}}{(2\pi)^{n-1}}\;
    \tilde{\phi}(\vec{k},\omega)
    \,e^{-i\vec{k}\vec{x}_0} &=& 0, \\
   (-\omega^2+|\vec{k}|^2+m^2)\;
    \tilde{\phi}(\vec{k},\omega)
    -\tilde{Q}(\omega)\,e^{i\vec{k}\vec{x}_0}&=&0.
\end{eqnarray}

After elimination of $\tilde{\phi}(\vec{k},\omega)$
from these equations,
one finds the effective equation of motion for the detector
\begin{equation}
\left(\widetilde{\frac{\delta S_0}{\delta Q}}\right)
+K_n(\omega)\,\tilde{Q}(\omega)=0,
\end{equation}
where
\begin{eqnarray}
 K_n(\omega)&=&\int \frac{d\vec{k}}{(2\pi)^{n-1}}\;
      \frac{1}{-\omega^2+|\vec{k}|^2+m^2\pm i\epsilon} \nonumber\\
  &=&\frac{2^{2-n}\pi^{\frac{1-n}{2}}}{\Gamma(\frac{n-1}{2})}
  \int_0^\infty 
  \frac{\kappa^{n-2}d\kappa}{-\omega^2+\kappa^2+m^2\pm i\epsilon}.
\label{TSK}
\end{eqnarray}
Note that we have added the $\pm i\epsilon$ terms to 
the denominator of the integrand to
avoid the singularity.
The sign is $+$ for $\omega>0$ and $-$ for $\omega<0$
due to causality.
(The integrand, as a function of $\omega$, 
must be analytic in the lower-half plane
of the complex $\omega$ plane.)
Thus, $K_n(\omega)$ has the imaginary part.
This gives the friction term.
We note that the conventional definition of
the friction term contains the time derivative of $Q(t)$,
which is converted to $i\omega$ by Fourier transformation.
So, the conventional definition of
the dissipative coefficient $R_n(\omega)$ is given by
\begin{equation}
{\rm Im}K_n(\omega)=-\omega R_n(\omega).
\end{equation}
The real part of $K_n(\omega)$ would
diverge for higher $n$.
But, fortunately, for our application of
the fluctuation-dissipation theorem,
we only need the imaginary part.
(The divergence of the real part of $K_n(\omega)$
would generally be renormalized by 
the potential of the detector.)

By using
\begin{equation}
   \frac{1}{x\pm i\epsilon}={\rm P}\,\frac{1}{x}\mp
   i\pi\delta(x)
\end{equation}
and
\begin{equation}
\delta(x^2-a^2)=\frac{1}{2|a|}
\left[\delta(x+a)+\delta(x-a)\right],
\end{equation}
one finds
\begin{equation}
 {\rm Im}K_n(\omega)=\mp
 \frac{2^{1-n}\pi^{\frac{3-n}{2}}}{\Gamma(\frac{n-1}{2})}
 \left(\sqrt{\omega^2-m^2} \right)^{n-3}\,\theta(\omega^2-m^2).
\end{equation}
Thus, the dissipative coefficient $R_n(\omega)$ is
\begin{equation}
  R_n(\omega)=\frac{1}{|\omega|}
  \frac{2^{1-n}\pi^{\frac{3-n}{2}}}{\Gamma(\frac{n-1}{2})}
 \left(\sqrt{\omega^2-m^2} \right)^{n-3}\,\theta(\omega^2-m^2).
\label{TSRm}
\end{equation}
Specially, for the massless case, $m=0$,
this becomes
\begin{equation}
  R_n(\omega)=
  \frac{2^{1-n}\pi^{\frac{3-n}{2}}}{\Gamma(\frac{n-1}{2})}
  |\omega|^{n-4},\qquad \mbox{(for $m=0$)}.
\label{TSR}
\end{equation}

By the fluctuation-dissipation theorem~\cite{CalWel51},
this dissipation suggests the fluctuation of the scalar field
if in thermal equilibrium.
The fluctuation of the scalar field is defined by
\begin{eqnarray}
    \langle\phi(\vec{x}_0,t)\phi(\vec{x}_0,t)\rangle_{\beta} & \equiv & 
  {\rm Tr}\left[e^{-\beta H}
  \phi(\vec{x}_0,t)\phi(\vec{x}_0,t)\right]
      /{\rm Tr}\left[e^{-\beta H}\right] \nonumber \\
     &=& \int dt' \langle\phi(\vec{x}_0,t)
        \phi(\vec{x}_0,t')\rangle_{\beta}\delta(t-t')\nonumber \\
  &=& \frac{1}{2\pi}\int^\infty_{-\infty} d\omega\; F_n(\omega),
\end{eqnarray}
where
\begin{equation}
  F_n(\omega)=\int dt'\;e^{-i\omega(t-t')}
  \langle\phi(\vec{x}_0,t)\phi(\vec{x}_0,t')\rangle_{\beta}
\end{equation}
is the power spectrum of the thermal noise.
Then, the fluctuation-dissipation theorem~\cite{CalWel51}
says that,
\begin{eqnarray}
 \int^\infty_{-\infty} d\omega\;F_n(\omega) 
 &=& 4 \int^\infty_0 d\omega\;
   \left[\frac{1}{2}\hbar\omega+
   \frac{\hbar\omega}{e^{\beta\hbar\omega}-1}\right]\;
   R_n(\omega) \nonumber \\
 &=& 2 \int^\infty_{-\infty} d\omega\;
   \frac{\hbar\omega R_n(|\omega|)}{e^{\beta\hbar\omega}-1}\\
 &=&\frac{2^{2-n}\pi^{\frac{3-n}{2}}}{\Gamma(\frac{n-1}{2})}
   \int^\infty_{-\infty} d\omega \;
   \frac{\hbar\omega|\omega|^{n-4}}{e^{\beta\hbar\omega}-1},
   \qquad \mbox{(for $m=0$)}.
\end{eqnarray}
This is perfectly consistent with the previous result.
(See, for example, Ref.\cite{Takagi86}.)

\subsection{Rindler Noise}
Next, we consider the Rindler noise.
The system again consists of a real scalar field
and a detector in $n$ dimensional flat spacetime.
However, the detector is now uniformly accelerated,
{\it i.e.}, the world line of the detector is
\begin{eqnarray}
 x^0(\tau)&=&a^{-1}\sinh a\tau, \\
 x^1(\tau)&=&a^{-1}\cosh a\tau, \\
 x^i(\tau)&=&{\rm const.},  \qquad\qquad (i=2,\cdots,n-1)
\end{eqnarray}
where $\tau$ is the proper time of the detector
and $a$ is the acceleration.
Therefore, it is convenient to take 
the Rindler coordinates~\cite{Rindle66}
\begin{eqnarray}
 x^0&=&\xi\sinh\eta, \\
 x^1&=&\xi\cosh\eta, \\
 x^i&=&x^i. \qquad\qquad (i=2,\cdots,n-1)
\end{eqnarray}
In these coordinates, the world line of the detector is
\begin{eqnarray}
 \eta(\tau)&=&a\tau, \\
 \xi(\tau)&=&a^{-1}, \\
 x^i(\tau)&=&{\rm const.}  \qquad\qquad (i=2,\cdots,n-1)
\end{eqnarray}
The detector is again linearly coupled with
the scalar field through 
an internal degree of freedom $Q(\tau)$, there.
However, the scalar field is initially in the ground state
in this case.
The action of the total system is
\begin{equation}
S = S_0(Q)+S_{int}(Q,\phi)+S_0(\phi),
\end{equation}
where
\begin{eqnarray}
S_0(Q)  &=& \int d\tau\;L(Q,\dot{Q}), \\
S_{int}(Q,\phi) &=& \int d\tau d\eta d\xi d\bmi{x}\;
               Q(\tau)\,\phi(\xi,\bmi{x},\eta)\;
               \delta(\xi-a^{-1})\,\delta(\eta-a\tau)\,
               \delta(\bmi{x}-\bmi{x}_0) \nonumber \\
        &=& \int d\tau\;Q(\tau)\,
               \phi(a^{-1},\bmi{x}_0,a\tau), \label{RSSint} \\
S_0(\phi) &=& \int d\eta d\xi d\bmi{x}\;\frac{1}{2} 
    \xi\,\left[ \frac{1}{\xi^2}(\partial_\eta\phi)^2
    -(\partial_\xi\phi)^2
  -\sum_{i=2}^{n-1}(\partial_i\phi)^2-m^2\phi^2 \right],\label{RSS0}
\end{eqnarray}
and $\bmi{x}_0$ is the position of the detector.
($\bmi{x}$ stands for an $(n-2)$-dimensional vector.)
$S_0(\phi)$ is the same as Eq.(\ref{TSS0})
but written in the Rindler coordinates.
Again, we do not need an explicit form of
$L(Q,\dot{Q})$ which is the Lagrangian for the (unperturbed) detector.

The equations of motion derived from this action are
\begin{eqnarray}
 \left(\frac{\delta S_0}{\delta Q}\right)
 +\int d\eta d\xi d\bmi{x}\;\phi(\xi,\bmi{x},\eta)
   \qquad\qquad\qquad\quad
   & & \nonumber \\
   {}\times\delta(\xi-a^{-1})\,\delta(\eta-a\tau)\,
   \delta(\bmi{x}-\bmi{x}_0) &=& 0
\end{eqnarray}
and
\begin{eqnarray}
\frac{1}{\xi^2}\partial_\eta^2\phi
   -\frac{1}{\xi}\partial_\xi(\xi\partial_\xi\phi)
   -\sum_i\partial_i^2\phi+m^2\phi
   \qquad\qquad\qquad
   & & \nonumber \\
  {}-\frac{1}{\xi}\int d\tau \;Q(\tau)\,
  \delta(\xi-a^{-1})\,\delta(\eta-a\tau)\,
  \delta(\bmi{x}-\bmi{x}_0)&=&0,
\end{eqnarray}
where
\begin{equation}
 -\left(\frac{\delta S_0}{\delta Q}\right)
 =\frac{d}{d\tau}\left(\frac{\partial L}{\partial \dot{Q}}\right)
 -\left(\frac{\partial L}{\partial Q}\right).
\end{equation}

In this case, we consider the transformations,
\begin{eqnarray}
\Phi(\nu,\bmi{k},\Omega) &\equiv& 
 \frac{1}{\pi} \int^{\infty}_0 \frac{d\xi}{\xi}
         \int d\eta d\bmi{x}\;
    \sqrt{2\nu\sinh\pi\nu}\,K_{i\nu}(M_k\xi)\nonumber \\
   & & \qquad\qquad\qquad\qquad\qquad 
      {}\times e^{i\bmi{k}\bmi{x}-i\Omega\eta}\,
     \phi(\xi,\bmi{x},\eta), \label{Phi} \\
\tilde{Q}(\omega) &\equiv&
   \int d\tau\;Q(\tau)\,e^{-i\omega \tau},
\end{eqnarray}
where
\begin{equation}
 M_k  \equiv \sqrt{m^2+|\bmi{k}|^2}
\end{equation}
and $K_{i\nu}(z)$ is a modified Bessel function of
imaginary order which satisfies
\begin{equation}
 \left\{ z^2\frac{d^2}{dz^2}+z\frac{d}{dz}
   -(z^2-\nu^2)\right\}\;K_{i\nu}(z)= 0.
\end{equation}
(Note that $\Omega$ is a frequency
with respect to the Rindler time $\eta$ and
$\omega$ is a frequency with respect to
the proper time of the detector $\tau$.)
By the orthogonality relation~\cite{Fullin73}
\begin{equation}
\frac{1}{\pi^2}\int_0^{\infty}\frac{dx}{x}\;
   K_{i\mu}(x)K_{i\nu}(x)
   =\frac{\delta(\mu-\nu)}{2\nu\sinh\pi\nu},
\label{SOR}
\end{equation}
we can write the inverse transformation of Eq.(\ref{Phi}) as
\begin{eqnarray}
 \phi(\xi,\bmi{x},\eta)  &=&
 \frac{1}{(2\pi)^{n-1}\pi} \int^{\infty}_0 d\nu 
   \int d\Omega d\bmi{k}\;
    \sqrt{2\nu\sinh\pi\nu}\,K_{i\nu}(M_k\xi) \nonumber \\
  & & \qquad\qquad\qquad\qquad\qquad\quad
    {}\times e^{-i\bmi{k}\bmi{x}+i\Omega\eta}\,
     \Phi(\nu,\bmi{k},\Omega).
\end{eqnarray}

Then, the equations of motion become
\begin{eqnarray}
 \left(\widetilde{\frac{\delta S_0}{\delta Q}}\right)
  +\frac{1}{(2\pi)^{n-2}\pi a}\int^{\infty}_0 d\nu 
    \int d\bmi{k}\;
    \sqrt{2\nu\sinh\pi\nu}\,K_{i\nu}(M_k/a) \nonumber & & \\
    {}\times e^{-i\bmi{k}\bmi{x}_0}\,
    \Phi(\nu,\bmi{k},\omega/a) &=& 0, \\
 (-\Omega^2+\nu^2)\,\Phi(\nu,\bmi{k},\Omega)
   -\frac{1}{\pi}\sqrt{2\nu\sinh\pi\nu}\,K_{i\nu}(M_k/a)
     \nonumber & & \\
    {}\times e^{i\bmi{k}\bmi{x}_0}\,\tilde{Q}(\Omega a)
    &=&0.
\end{eqnarray}

After elimination of $\Phi(\nu,\bmi{k},\Omega)$
from these equations,
one finds the effective equation of motion for the detector
\begin{equation}
\left(\widetilde{\frac{\delta S_0}{\delta Q}}\right)
+{\cal K}_n(\omega)\,\tilde{Q}(\omega)=0,
\end{equation}
where
\begin{eqnarray}
 {\cal K}_n(\omega)&=&\frac{1}{2^{n-2}\pi^n a}
   \int^\infty_0 d\nu \int d\bmi{k}\;
  \frac{2\nu\sinh\pi\nu}{-(\omega/a)^2+\nu^2\pm i\epsilon}\,
  \left[K_{i\nu}(M_k/a)\right]^2 \nonumber \\
  &=&\frac{2^{3-n}}{\pi^{\frac{n+2}{2}}\Gamma(\frac{n-2}{2})\;a}
    \int^\infty_0  d\nu
    \frac{2\nu\sinh\pi\nu}{-(\omega/a)^2+\nu^2\pm i\epsilon}
    \nonumber \\
  & &\qquad\qquad\qquad\qquad
    {}\times \int^\infty_0 d\kappa \;\kappa^{n-3}
    \left[K_{i\nu}(M_\kappa/a)\right]^2.
\label{RSK}
\end{eqnarray}
Again, we have added the $\pm i\epsilon$ term to 
the denominator of the integrand.
Specially, for the massless case, $m=0$,
we can use a closed form of integration given
by the formula~\cite{GraRyz80}
\begin{eqnarray}
  \int^\infty_0dx\;x^{n-3}\left[K_\mu(x)\right]^2
  &=&\frac{2^{n-5}}{\Gamma(n-2)}
    \left[\Gamma\left(\frac{n-2}{2}\right)\right]^2
    \nonumber \\
  & & {}\times
    \Gamma\left(\frac{n}{2}-1+\mu\right)
    \Gamma\left(\frac{n}{2}-1-\mu\right).
\end{eqnarray}
Thus, one obtains for Eq.(\ref{RSK})
\begin{equation}
 {\cal K}_n(\omega)=\frac{a^{n-3}\Gamma(\frac{n-2}{2})}
     {2\pi^{\frac{n}{2}}\Gamma(n-2)}
    \int^\infty_0 d\nu 
    \frac{1}{-(\omega/a)^2+\nu^2\pm i\epsilon}\,
    \left|\frac{\Gamma(\frac{n}{2}-1+i\nu)}{\Gamma(i\nu)}\right|^2,
\end{equation}
where we have used
\begin{equation}
|\Gamma(iy)|^2=\frac{\pi}{y\sinh\pi y}  
  \qquad (y:{\rm real}).
\label{Sinh}
\end{equation}
(Although $\kappa$ integral is absent for $n=2$,
this expression is also valid for $n=2$.)
By using,
\begin{equation}
\Gamma(2z)=\frac{2^{2z}}{2\sqrt{\pi}}\;\Gamma(z)\;
\Gamma\left(z+\frac{1}{2}\right),
\end{equation}
one finds that the dissipative coefficient ${\cal R}_n(\omega)$ is
\begin{eqnarray}
 {\cal R}_n(\omega) &\equiv& 
     \frac{{\rm Im}{\cal K}_n(\omega)}{-\omega} \nonumber \\
    &=& \frac{2^{1-n}\pi^{\frac{3-n}{2}}}{\Gamma(\frac{n-1}{2})}
        \frac{a^{n-2}}{\omega^2}
 \left|\frac{\Gamma(\frac{n}{2}-1+i\frac{\omega}{a})}
         {\Gamma(i\frac{\omega}{a})}\right|^2, 
       \qquad \mbox{(for $m=0$)}
\label{RSR}
\end{eqnarray}
and that the ratio of this Rindler case to
the thermal case Eq.(\ref{TSR}) is given by
\begin{eqnarray}
  r_n(\omega)&=&{\cal R}_n(\omega)\,/\,R_n(\omega) \\
           &=& \left|\frac{\omega}{a}\right|^{2-n}
 \left|\frac{\Gamma(\frac{n}{2}-1+i\frac{\omega}{a})}
         {\Gamma(i\frac{\omega}{a})}\right|^2.
\label{Ratio}
\end{eqnarray}

By using the elementary formulas
\begin{eqnarray}
  \Gamma(z+1)&=&z\,\Gamma(z), \\
  \Gamma\left(\frac{1}{2}+z\right)\,
  \Gamma\left(\frac{1}{2}-z\right)
    &=&\frac{\pi}{\cos\pi z},
\end{eqnarray}
and Eq.(\ref{Sinh}),
one finds, for even $n$, 
\begin{eqnarray}
 \left|\Gamma\left(\frac{n}{2}-1+
     i\frac{\omega}{a}\right)\right|^2&=&
 \left[\left(\frac{n}{2}-2\right)^2+
        \left(\frac{\omega}{a}\right)^2\right]
 \left[\left(\frac{n}{2}-3\right)^2+
        \left(\frac{\omega}{a}\right)^2\right]
 \cdots  \nonumber \\
 & & \cdots
 \left[1^2+\left(\frac{\omega}{a}\right)^2\right]
   \left(\frac{\omega}{a}\right)^2
 \left|\Gamma\left(i\frac{\omega}{a}\right)\right|^2 
         \nonumber \\
 &\equiv& d_n(\omega)\left|\frac{\omega}{a}\right|^{n-2}
   \left|\Gamma\left(i\frac{\omega}{a}\right)\right|^2,
\end{eqnarray}
and, for odd $n$,
\begin{eqnarray}
 \left|\Gamma\left(\frac{n}{2}-1+
     i\frac{\omega}{a}\right)\right|^2&=&
 \left[\left(\frac{n}{2}-2\right)^2+
        \left(\frac{\omega}{a}\right)^2\right]
 \left[\left(\frac{n}{2}-3\right)^2+
        \left(\frac{\omega}{a}\right)^2\right]
 \cdots  \nonumber \\
 & & \cdots
 \left[\left(\frac{1}{2}\right)^2+
        \left(\frac{\omega}{a}\right)^2\right]
 \left|\Gamma\left(\frac{1}{2}+i\frac{\omega}{a}\right)\right|^2
          \nonumber \\
 &\equiv& d_n(\omega)\left|\frac{\omega}{a}\right|^{n-2}
          \left(\frac{\omega}{a}\right)^{-1}
          \left|\Gamma\left(\frac{1}{2}+
             i\frac{\omega}{a}\right)\right|^2.
\end{eqnarray}
Then, the ratio in Eq.(\ref{Ratio}) can be written as
\begin{equation}
r_n(\omega)=\cases{
    d_n(\omega) & ($n$:even) \cr
    \tanh(\pi\omega/a)\;
    d_n(\omega) & ($n$:odd)  \cr }.
\end{equation}

Note that
\begin{equation}
    \lim_{a\to0}r_n(\omega)=1,
\end{equation}
that is,
\begin{equation}
    \lim_{a\to0}{\cal R}_n(\omega)=R_n(\omega).
\end{equation}

By the fluctuation-dissipation theorem,
this dissipation means the fluctuation of the scalar field
if in ``thermal'' equilibrium.
Although the scalar field is in the vacuum state in this case,
we can view it as in thermal equilibrium 
at temperature $T=\hbar a/2\pi$
by the ``thermalization theorem''.
There are various versions of the theorem.
(For a review, see Ref.\cite{Takagi86}.)

For example, since
the positive Rindler wedge $R_+$
(a quarter of the Minkowski spacetime, $x^1>|x^0|$)
is causally disconnected from
the negative Rindler wedge $R_-$
(another quarter of the Minkowski spacetime, $x^1<-|x^0|$),
a uniformly accelerated observer who will be permanently
confined within $R_+$ does not concern with 
the degrees of the freedom associated with $R_-$.
Then, by tracing out over these degrees 
from the Minkowski vacuum, the observer gets
the thermal density matrix
at temperature $T=\hbar a/2\pi$~\cite{UnrWal84}.
More specifically~\cite{Israel76,TroVan79,Takagi86},
\begin{equation}
    \langle0_M|O^{(+)}|0_M\rangle=
     {\rm Tr}\left[e^{-\frac{2\pi}{a}aH_R^{(+)}}O^{(+)}\right]
   /{\rm Tr}\left[e^{-\frac{2\pi}{a}aH_R^{(+)}}\right],
\label{thermal}
\end{equation}
where $O^{(+)}$ is an operator only for $R_+$
and $H_R^{(+)}$ is the Rindler Hamiltonian 
(the generator of $\eta$ translation,
{\it i.e.}, the boost) restricted to there.
Note that, for the uniformly accelerated observer,
the generator of $\tau$ translation is $aH_R^{(+)}$.
That is,
as far as the uniformly accelerated observer is concerned,
the expectation value in the Minkowski vacuum
of the operator can be seen as an ensemble average over
the density matrix which is the same form as canonical ensemble
at temperature $T=\hbar a/2\pi$.
This effective canonical ensemble is true even
for interacting fields~\cite{UnrWei84,HoHoYa85}.

Equivalently, one can see the thermal character from
the periodicity of the propagator
in the imaginary time~\cite{ChrDuf78,TroVan79}.
Moreover, it was shown by Sewell~\cite{Sewell82} 
in the context of the axiomatic quantum field theory that
the Rindler noise satisfies
the KMS condition~\cite{Kubo57,MarSch59}
(which is the definition of the thermal equilibrium
with the systems with infinite numbers of degrees of freedom)
at temperature $T=\hbar a/2\pi$
for general interacting field of any spin in any dimension.
The intuitive explanation of Sewell's theorem is found in
Ref.\cite{Ooguri86}.

From these various versions of the 
thermalization theorem, 
one might conclude without calculations that
the Rindler noise is the same as the thermal noise 
at temperature $T=\hbar a/2\pi$.
However, these results {\em only} mean that the
uniformly accelerated observer would see the Minkowski vacuum
as the thermal equilibrium state
but do {\em not} mean that the
uniformly accelerated detector would respond in
the same way as it would do at rest in the thermal bath,
as was emphasized by Unruh-Wald~\cite{UnrWal84} 
and Takagi~\cite{Takagi86}.
The thermalization theorem, itself,
says nothing about the noise.
In fact, the Rindler noise is different from
the thermal noise in several points.

Now, we shall calculate the Rindler noise.
By using Eq.(\ref{RSR}), the thermalization theorem and
the fluctuation-dissipation theorem,
we can take a shortcut to calculate it.
The fluctuation of the scalar field is defined by
\begin{equation}
    \langle0_M|\phi(\tau)\phi(\tau)|0_M\rangle
    =\frac{1}{2\pi}\int^\infty_{-\infty}d\omega\;{\cal F}_n(\omega),
\end{equation}
where $\phi(\tau)=\phi(\xi(\tau),\bmi{x}(\tau),\eta(\tau))$
is the value of the field along the world line
of the detector and
\begin{equation}
  {\cal F}_n(\omega)=\int d\tau'\;e^{-i\omega(\tau-\tau')}
  \langle0_M|\phi(\tau)\phi(\tau')|0_M\rangle
\end{equation}
is the power spectrum of the Rindler noise.
Then, by using Eq.(\ref{thermal}),
the fluctuation-dissipation theorem says that,
for the massless case,
\begin{eqnarray}
 \int^\infty_{-\infty} d\omega\;{\cal F}_n(\omega) 
 &=& 4 \int^\infty_0 d\omega\;
   \left[\frac{1}{2}\hbar\omega+
   \frac{\hbar\omega}{e^{\beta\hbar\omega}-1}\right]\;
   {\cal R}_n(\omega) \nonumber \\
 &=& 2 \int^\infty_{-\infty} d\omega\;
   \frac{\hbar\omega {\cal R}_n(|\omega|)}{e^{\beta\hbar\omega}-1}
   \nonumber \\
 &=& 2 \int^\infty_{-\infty} d\omega\;
   \frac{\hbar\omega R_n(|\omega|)}{e^{\beta\hbar\omega}-1}\;
   r_n(|\omega|) \nonumber \\
 &=& 2 \int^\infty_{-\infty} d\omega\;
   \frac{\hbar\omega R_n(|\omega|)}{e^{\beta\hbar\omega}-(-1)^{n}}
   \;d_n(\omega),\quad \mbox{(for $m=0$)},
\label{RSF}
\end{eqnarray}
where
\begin{eqnarray}
 d_2(\omega)  &=& 1, \\
 d_3(\omega)  &=& 1\times \frac{\omega}{|\omega|}, \\
 d_4(\omega)  &=& 1, \\
 d_5(\omega)  &=& \left[1+\left(\frac{a}{2\omega}
                  \right)^2\right]
                  \times\frac{\omega}{|\omega|}, \\
 d_6(\omega)  &=& \left[1+\left(\frac{a}{\omega}
                  \right)^2\right].
\end{eqnarray}
This result including the statistical inversion in the denominator
of Eq.(\ref{RSF}) is consistent with that of Takagi~\cite{Takagi86},
except for a minor difference in
$d_n(\omega)$ for odd $n$.
(We believe that our result is a correct one.)
Note that, in Ref.\cite{Takagi86},
the case of the complex scalar field was calculated,
but the result is the same for the real scalar field
as will be explained in the next section.

We have derived Eq.(\ref{RSF}) on the basis of Eq.(\ref{thermal})
and the fluctuation-dissipation theorem of a bosonic operator.
This clearly shows that the statistics inversion
is an ``apparent'' one not based on the basic principle
of statistical mechanics but based on the ``temperature''
dependence of the dissipative coefficient.

%% file: complex.tex
Now, we generalize the analysis to that of a complex scalar field.
The situation is the same as the real scalar case,
except that $Q$ and $\phi$ become
non-hermitian.

\subsection{Thermal Noise}
For the thermal noise,
Eqs.(\ref{TSSint}) and (\ref{TSS0})
are replaced by
\begin{eqnarray}
S_{int}(Q,\phi)&=&\int dtd\vec{x}\;
    \left[Q(t)\,\phi(\vec{x},t)+
        Q^\dagger(t)\,\phi^\dagger(\vec{x},t)\right]\;
               \delta(\vec{x}-\vec{x}_0) \nonumber \\
        &=& \int dt\;
    \left[Q(t)\,\phi(\vec{x}_0,t)+
        Q^\dagger(t)\,\phi^\dagger(\vec{x}_0,t)\right], \\
S_0(\phi) &=& \int dtd\vec{x}\;
  \left[ |\partial_t\phi|^2-\sum_{i=1}^{n-1}
         |\partial_i\phi|^2-m^2|\phi|^2 \right].
\end{eqnarray}
Thus, the effective equation of motion for the detector is
\begin{equation}
\left(\widetilde{\frac{\delta S_0}{\delta Q^\dagger}}\right)
+K_n(\omega)\,\tilde{Q}(\omega)=0,
\end{equation}
where $K_n(\omega)$ is the same function as
the real scalar case Eq.(\ref{TSK}).
Therefore, the dissipative coefficient is the {\em same} as
the real scalar case Eq.(\ref{TSR}).

Then, we can use the fluctuation-dissipation theorem
of a non-hermitian operator~\cite{FujTer98}.
The fluctuation of the field is now defined by
\begin{equation}
    \langle\phi(\vec{x}_0,t)\phi^\dagger(\vec{x}_0,t)\rangle_{\beta}= 
   \frac{1}{2\pi}\int^\infty_{-\infty} d\omega\; F_n(\omega),
\end{equation}
where
\begin{equation}
  F_n(\omega)=\int dt'\;e^{-i\omega(t-t')}
  \langle\phi(\vec{x}_0,t)\phi^\dagger(\vec{x}_0,t')\rangle_{\beta},
\end{equation}
and the theorem says that
\begin{eqnarray}
 \int^\infty_{-\infty} d\omega\;F_n(\omega) 
 &=& 4 \int^\infty_0 d\omega\;
   \frac{\hbar\omega}{2}\Bigl\{
   \left[1+\frac{1}{e^{\beta\hbar\omega}-1}\right]\,R_n(\omega)
    \nonumber \\ & & \qquad\qquad\qquad\qquad  
   {}+\left[\frac{1}{e^{\beta\hbar\omega}-1}\right]\,R_n(-\omega)
   \Bigr\} \nonumber \\
 &=& 2 \int^\infty_{-\infty} d\omega\;
   \frac{\hbar\omega R_n(-\omega)}{e^{\beta\hbar\omega}-1} \\
 &=&\frac{2^{2-n}\pi^{\frac{3-n}{2}}}{\Gamma(\frac{n-1}{2})}
   \int^\infty_{-\infty} d\omega \;
   \frac{\hbar\omega|\omega|^{n-4}}{e^{\beta\hbar\omega}-1}.
   \qquad \mbox{(for $m=0$)}
\end{eqnarray}

\subsection{Rindler Noise}
Similarly, for the Rindler noise,
Eqs.(\ref{RSSint}) and (\ref{RSS0})
are replaced by
\begin{eqnarray}
S_{int}(Q,\phi) &=& \int d\tau d\eta d\xi d\bmi{x}\;
    \left[Q(\tau)\,\phi(\xi,\bmi{x},\eta)+
        Q^\dagger(\tau)\,\phi^\dagger(\xi,\bmi{x},\eta)\right]
        \nonumber \\ & & \qquad\qquad\qquad\qquad {}\times
               \delta(\xi-a^{-1})\,\delta(\eta-a\tau)\,
               \delta(\bmi{x}-\bmi{x}_0) \nonumber \\
        &=& \int d\tau\;
    \left[Q(\tau)\,\phi(a^{-1},\bmi{x}_0,a\tau)+
   Q^\dagger(\tau)\,\phi^\dagger(a^{-1},\bmi{x}_0,a\tau)\right], \\
S_0(\phi) &=& \int d\eta d\xi d\bmi{x}\,
    \xi\left[ \frac{1}{\xi^2}|\partial_\eta\phi|^2
    -|\partial_\xi\phi|^2
    -\sum_{i=2}^{n-1}|\partial_i\phi|^2-m^2|\phi|^2 \right],
\end{eqnarray}
and the effective equation of motion becomes
\begin{equation}
\left(\widetilde{\frac{\delta S_0}{\delta Q^\dagger}}\right)
+{\cal K}_n(\omega)\,\tilde{Q}(\omega)=0,
\end{equation}
where ${\cal K}_n(\omega)$ is the same function as
the real scalar case Eq.(\ref{RSK}).
So, the dissipative coefficient is the {\em same} as
the real scalar case Eq.(\ref{RSR}), again.

The fluctuation of the field is now defined by
\begin{equation}
    \langle0_M|\phi(\tau)\phi^\dagger(\tau)|0_M\rangle
    =\frac{1}{2\pi}\int^\infty_{-\infty}d\omega\;{\cal F}_n(\omega),
\end{equation}
where
\begin{equation}
  {\cal F}_n(\omega)=\int d\tau'\;e^{-i\omega(\tau-\tau')}
  \langle0_M|\phi(\tau)\phi^\dagger(\tau')|0_M\rangle,
\end{equation}
and, by using the fluctuation-dissipation theorem
of a non-hermitian operator~\cite{FujTer98} and Eq.(\ref{thermal}), 
one finds that
\begin{eqnarray}
 \int^\infty_{-\infty} d\omega\;{\cal F}_n(\omega) 
 &=& 4 \int^\infty_0 d\omega\;
   \frac{\hbar\omega}{2}\Bigl\{
   \left[1+\frac{1}{e^{\beta\hbar\omega}-1}\right]\,
    {\cal R}_n(\omega)
    \nonumber \\ & & \qquad\qquad\qquad\qquad  
   {}+\left[\frac{1}{e^{\beta\hbar\omega}-1}\right]\,
    {\cal R}_n(-\omega)
    \Bigr\} \nonumber \\
 &=& 2 \int^\infty_{-\infty} d\omega\;
   \frac{\hbar\omega {\cal R}_n(-\omega)}{e^{\beta\hbar\omega}-1}
   \\
 &=& 2 \int^\infty_{-\infty} d\omega\;
   \frac{\hbar\omega R_n(-\omega)}{e^{\beta\hbar\omega}-(-1)^{n}}\;
   d_n(\omega).\quad \mbox{(for $m=0$)}
\end{eqnarray}
in agreement with the previous analysis~\cite{Takagi86}.

%% file: dirac.tex
Next, we generalize to a Dirac field,
using the fluctuation-dissipation theorem
of a fermionic operator~\cite{FujTer98}.
The situation is the same as the scalar case.
However, the internal degree of freedom of the detector
is now a spinor $\Theta$ and the field is
a Dirac field $\psi$.

\subsection{Thermal Noise}
For the thermal noise,
the action of the total system is
\begin{equation}
S = S_0(\Theta)+S_{int}(\Theta,\psi)+S_0(\psi),
\end{equation}
where
\begin{eqnarray}
S_0(\Theta)  &=& \int dt\;L(\Theta,\dot{\Theta}), \\
S_{int}(\Theta,\psi)&=&\int dtd\vec{x}\
      \left[\bar{\Theta}(t)\,\psi(\vec{x},t)+
        \bar{\psi}(\vec{x},t)\,\Theta(t)\right]\;
               \delta(\vec{x}-\vec{x}_0) \nonumber \\
        &=& \int dt\;
      \left[\bar{\Theta}(t)\,\psi(\vec{x}_0,t)+
        \bar{\psi}(\vec{x}_0,t)\,\Theta(t)\right], \\
S_0(\psi) &=& \int dtd\vec{x}\;\,\bar{\psi}
  \left[ i\gamma^\mu\partial_\mu-m \right]\psi,
\label{TDS0}
\end{eqnarray}
and
\begin{equation}
  \bar{\psi}\equiv\psi^\dagger\gamma^0,\qquad
  \bar{\Theta}\equiv\Theta^\dagger\gamma^0.
\end{equation}
Here, $\gamma^\mu$ are the $\gamma$ matrices
which satisfy
\begin{equation}
  \{ \gamma^\mu,\gamma^\nu \}=-2\eta^{\mu\nu}.
\end{equation}

The equations of motion become
\begin{eqnarray}
  \left(\frac{\delta_L S_0}{\delta\bar{\Theta}}\right)
 +\int d\vec{x}\;\psi(\vec{x},t)
    \,\delta(\vec{x}-\vec{x}_0) &=& 0, \\
\partial_t^2\psi-\sum_i\partial_i^2\psi+m^2\psi 
 - \left[ i\gamma^\mu\partial_\mu+m \right]
 \Theta(t)\,\delta(\vec{x}-\vec{x}_0)&=&0,
\end{eqnarray}
where ``L'' denotes the left-derivative.
After Fourier transformation and elimination of the field,
one finds that the effective equation of motion
for the detector is
\begin{equation}
\left(\widetilde{\frac{\delta_L S_0}{\delta\bar{\Theta}}}\right)
+K_{1/2,n}(\omega)\,\tilde{\Theta}(\omega)=0,
\end{equation}
where
\begin{equation}
 K_{1/2,n}(\omega)=\int \frac{d\vec{k}}{(2\pi)^{n-1}}\;
      \frac{-\omega\gamma^0+\vec{k}\cdot\vec{\gamma}+m}%
     {-\omega^2+|\vec{k}|^2+m^2\pm i\epsilon}.
\end{equation}
(The subscript ``$1/2$'' denotes the spin of field.)
Since the $\vec{k}\cdot\vec{\gamma}$ term is odd in $\vec{k}$,
we can relate $K_{1/2,n}(\omega)$ to that of the scalar case
Eq.(\ref{TSK}),
\begin{equation}
  K_{1/2,n}(\omega)=
  \left[ -\omega\gamma^0+m \right]\;K_n(\omega).
\end{equation}

Therefore, we define the dissipative coefficient as
\begin{eqnarray}
 R_{1/2,n}(\omega)&\equiv&
     -\frac{{\rm Im}K_{1/2,n}(\omega)}{-\omega} \nonumber \\
  &=&-\left[-\omega\gamma^0+m\right]\;R_n(\omega),
\end{eqnarray}
where $R_n(\omega)$ is that of the scalar case Eq.(\ref{TSRm}).
This extra $-$ sign is a convention; it was chosen so that
the dissipative coefficient is ``positive''
in the sense that
\begin{equation}
    {\rm Tr}\;\gamma^0R_{1/2,n}(\omega)\propto
    \omega R_n(\omega)\ge0
   \qquad \mbox{for $\omega\ge0$},
\end{equation}
and this convention is the same as in Ref.\cite{FujTer98}.
We also define
\begin{equation}
 \bar{R}_{1/2,n}(-\omega)\equiv-R_{1/2,n}(-\omega)=
    \left[\omega\gamma^0+m\right]\,R_n(-\omega)
\end{equation}
and
\begin{equation}
    {\rm Tr}\;\gamma^0\bar{R}_{1/2,n}(-\omega)\propto
    \omega R_n(-\omega)\ge0
   \qquad \mbox{for $\omega\ge0$}.
\end{equation}

Then, we can use the fluctuation-dissipation theorem
of a fermionic operator~\cite{FujTer98}.
The thermal fluctuation of the Dirac field is defined by
\begin{eqnarray}
\langle\psi(\vec{x}_0,t)\bar{\psi}(\vec{x}_0,t)\rangle_{\beta}
 & \equiv & {\rm Tr}\left[e^{-\beta H}
  \psi(\vec{x}_0,t)\bar{\psi}(\vec{x}_0,t)\right]
      /{\rm Tr}\left[e^{-\beta H}\right] \nonumber \\
     &=& \int dt' \langle\psi(\vec{x}_0,t)
        \bar{\psi}(\vec{x}_0,t')\rangle_{\beta}\delta(t-t')
        \nonumber \\
     &=& \frac{1}{2\pi}\int dt'd\omega\;
     e^{-i\omega(t-t')} \nonumber \\
  & & \qquad\qquad {}\times
  \langle\psi(\vec{x}_0,t)\bar{\psi}(\vec{x}_0,t')\rangle_{\beta}.
\end{eqnarray}
The power spectrum of the Dirac field is defined by
\begin{equation}
  F_{1/2,n}(\omega)\equiv\Delta_n^{-1} {\rm Tr}\left[
    \gamma^0\int dt'\;e^{-i\omega(t-t')}
  \langle\psi(\vec{x}_0,t)\bar{\psi}(\vec{x}_0,t')\rangle_{\beta}
   \right],
\end{equation}
where $\Delta_n$ is the dimension of the $\gamma$ matrices.
Therefore,
\begin{equation}
{\rm Tr}\left[\gamma^0
\langle\psi(\vec{x}_0,t)\bar{\psi}(\vec{x}_0,t)\rangle_{\beta}
\right]
=\frac{\Delta_n}{2\pi}\int_{-\infty}^\infty d\omega\;
F_{1/2,n}(\omega).
\end{equation}

Then, the theorem~\cite{FujTer98} says that
\begin{eqnarray}
 \int^\infty_{-\infty} d\omega\;F_{1/2,n}(\omega) 
 &=& \frac{4}{\Delta_n} {\rm Tr}\Biggl[\gamma^0
    \int^\infty_0 d\omega\;
   \frac{\hbar\omega}{2}\Bigl\{
   \left[1-\frac{1}{e^{\beta\hbar\omega}+1}\right]\,R_{1/2,n}(\omega)
    \nonumber \\ & & \qquad\qquad\qquad\qquad  
   {}+\left[\frac{1}{e^{\beta\hbar\omega}+1}\right]\,
    \bar{R}_{1/2,n}(-\omega) \Bigr\} \Biggr]\nonumber \\
 &=& \frac{2}{\Delta_n} \int^\infty_{-\infty} d\omega\;
   \frac{\hbar\omega}{e^{\beta\hbar\omega}+1}
    {\rm Tr}[\gamma^0\bar{R}_{1/2,n}(-\omega)] \nonumber \\
 &=& 2\int^\infty_{-\infty} d\omega\;
   \frac{\hbar\omega^2R_n(-\omega)}{e^{\beta\hbar\omega}+1}.
\end{eqnarray}
This result is consistent with the previous result.
(See, for example, Ref.\cite{Takagi86}.)
The differences between this case and the scalar case
are the power of $\omega$ in the numerator,
which comes from the fact that the Dirac equation is 
the first order with respect to the time derivative,
and the Fermi distribution factor,
which is provided by the fluctuation-dissipation theorem
of a fermionic operator.

\subsection{Rindler Noise}
For the Rindler noise,
one might define the Rindler noise of Dirac field as
\begin{equation}
   \langle0_M|\psi(\tau)\bar{\psi}(\tau')|0_M\rangle.
\end{equation}
However, this noise is not stationary,
{\it i.e.}, is not a function of $\tau-\tau'$.
Instead, we define the Rindler noise of 
Dirac field as~\cite{Takagi86,TroVan79}
\begin{equation}
   \langle0_M|\hat{\psi}(\tau)\bar{\hat{\psi}}(\tau')|0_M\rangle,
\end{equation}
where
\begin{eqnarray}
   \hat{\psi}(\tau)&=&
    \exp\left[-\frac{1}{2}a\tau\gamma^0\gamma^1\right]
     \psi(\tau) \nonumber \\
    &=& \left[\cosh(a\tau/2)-\gamma^0\gamma^1
       \sinh(a\tau/2)\right]\psi(\tau) \nonumber \\
    &\equiv& S_\tau\psi(\tau).
\end{eqnarray}
This transformation is the Lorentz transformation, the boost,
from the laboratory frame to the instantaneously comoving frame
of the uniformly accelerated observer at $\tau$.

Thus, we write the action in terms of
\begin{eqnarray}
  \hat{\psi}(\xi,\bmi{x},\eta)&\equiv& 
    S_{\frac{\eta}{a}}\psi(\xi,\bmi{x},\eta),  \\
  \hat{\Theta}(\tau) &\equiv&
    S_\tau\Theta(\tau).
\end{eqnarray}
Then, the action of the total system is
\begin{equation}
S = S_0(\hat{\Theta})+S_{int}(\hat{\Theta},\hat{\psi})
    +S_0(\hat{\psi}),
\end{equation}
where
\begin{eqnarray}
S_0(\hat{\Theta})  &=& \int 
       d\tau\;L(\hat{\Theta},\dot{\hat{\Theta}}), \\
S_{int}(\hat{\Theta},\hat{\psi})&=&\int d\tau d\eta d\xi d\bmi{x}\;
      \left[\bar{\hat{\Theta}}(\tau)\,
         \hat{\psi}(\xi,\bmi{x},\eta)+
         \bar{\hat{\psi}}(\xi,\bmi{x},\eta)\,
         \hat{\Theta}(\tau)\right]\;\nonumber \\ 
& & \qquad\qquad\qquad {}\times
         \delta(\xi-a^{-1})\,\delta(\eta-a\tau)\,
         \delta(\bmi{x}-\bmi{x}_0), \\
S_0(\hat{\psi}) &=& \int d\eta d\xi d\bmi{x}\;\xi
        \,\bar{\hat{\psi}}
  \Bigl[ i\gamma^0\frac{1}{\xi}\partial_\eta
        +i\gamma^1\left(\frac{1}{2\xi}+\partial_\xi\right)
 \nonumber \\ 
 & & \qquad\qquad\qquad\qquad\qquad
  {}+i\sum_{i=2}^{n-1}\gamma^i\partial_i-m \Bigr]\hat{\psi}.
\end{eqnarray}
This $S_0(\hat{\psi})$ is the same as Eq.(\ref{TDS0})
but written in the Rindler coordinates and $\hat{\psi}$.
(If we introduce the vielbein,
calculate the spin connection in the Rindler space and 
define the spinor with respect to the local Lorentz,
we would get this action.)

The equations of motion become
\begin{eqnarray}
 \left(\frac{\delta_L S_0}{\delta\bar{\hat{\Theta}}}\right)
 +\int d\eta d\xi d\bmi{x}\;\hat{\psi}(\xi,\bmi{x},\eta)
   \qquad\qquad\qquad\quad
   & & \nonumber \\
   {}\times\delta(\xi-a^{-1})\,\delta(\eta-a\tau)\,
   \delta(\bmi{x}-\bmi{x}_0) &=& 0,
\end{eqnarray}
and
\begin{eqnarray}
\frac{1}{\xi^2}\partial_\eta^2\hat{\psi}
  +\frac{1}{\xi^2}\gamma^0\gamma^1\partial_\eta\hat{\psi}
  +\frac{1}{4\xi^2}\hat{\psi}
-\frac{1}{\xi}\partial_\xi(\xi\partial_\xi\hat{\psi})
   -\sum_i\partial_i^2\hat{\psi}+m^2\hat{\psi}
   & & \nonumber \\
  {}-\left[ i\gamma^0\frac{1}{\xi}\partial_\eta
   +i\gamma^1\left(\frac{1}{2\xi}+\partial_\xi\right)
   +i\sum_i\gamma^i\partial_i+m \right]
   & & \nonumber \\
{}\times\frac{1}{\xi}\int d\tau \;\hat{\Theta}(\tau)\,
  \delta(\xi-a^{-1})\,\delta(\eta-a\tau)\,
  \delta(\bmi{x}-\bmi{x}_0)&=&0.
\end{eqnarray}

In order to deal with the $\gamma^0\gamma^1$ term of 
the second equation,
we define the projection operators
\begin{equation}
   \gamma_{\pm}\equiv\left(\frac{1\pm\gamma^0\gamma^1}{2}\right),
\end{equation}
which satisfy
\begin{equation}
  \gamma_{\pm}^2=\gamma_{\pm}, \qquad
  \gamma_{+}\gamma_{-}=0,      \qquad
  \gamma_{+}+\gamma_{-}=1,     \qquad
  \gamma_{\pm}(\gamma^0\gamma^1)=\pm\gamma_{\pm}.
\end{equation}
By using these projection operators,
we can divide $\hat{\psi}$ into two parts
\begin{equation}
   \hat{\psi}=\hat{\psi}_++\hat{\psi}_-,
\end{equation}
where $\hat{\psi}_\pm\equiv\gamma_{\pm}\hat{\psi}$.
Then, the second equation becomes
\begin{eqnarray}
\frac{1}{\xi^2}\left(\partial_\eta\pm\frac{1}{2}\right)^2
\hat{\psi}_\pm
-\frac{1}{\xi}\partial_\xi(\xi\partial_\xi\hat{\psi}_\pm)
   -\sum_i\partial_i^2\hat{\psi}_\pm+m^2\hat{\psi}_\pm
   & & \nonumber \\
  {}-\gamma_\pm\left[ i\gamma^0\frac{1}{\xi}\partial_\eta
   +i\gamma^1\left(\frac{1}{2\xi}+\partial_\xi\right)
   +i\sum_i\gamma^i\partial_i+m \right]
   & & \nonumber \\
{}\times\frac{1}{\xi}\int d\tau \;\hat{\Theta}(\tau)\,
  \delta(\xi-a^{-1})\,\delta(\eta-a\tau)\,
  \delta(\bmi{x}-\bmi{x}_0)&=&0.
\end{eqnarray}

In this case, instead of $K_{i\nu}(M_k\xi)$,
we consider $K_{i\nu\pm\frac{1}{2}}(M_k\xi)$ as
\begin{eqnarray}
\hat{\Psi}_\pm(\nu,\bmi{k},\Omega) &\equiv& 
 \frac{1}{\pi} \int^{\infty}_0 \frac{d\xi}{\xi}
         \int d\eta d\bmi{x}\;
    K_{i\nu\pm\frac{1}{2}}(M_k\xi)\nonumber \\
   & & \qquad\qquad\qquad\qquad 
      {}\times e^{i\bmi{k}\bmi{x}-i\Omega\eta}\,
     \hat{\psi}_\pm (\xi,\bmi{x},\eta).
\label{Psi}
\end{eqnarray}
(See, for example, Ref.\cite{SoMuGr80}.)
By the orthogonality relation
\begin{equation}
\frac{1}{\pi^2}\int_0^{\infty}\frac{dx}{x}\;
   K_{i\mu\pm\frac{1}{2}}(x)K_{i\nu\pm\frac{1}{2}}(x)
   =\frac{\delta(\mu-\nu)}{(\mp 2i\nu-1)\cosh\pi\nu},
\end{equation}
which can be obtained from Eq.(\ref{SOR})
by the shift $i\nu\to i\nu\pm\frac{1}{2}$,
we can invert Eq.(\ref{Psi}) as
\begin{eqnarray}
 \hat{\psi}_\pm(\xi,\bmi{x},\eta)  &=&
 \frac{1}{(2\pi)^{n-1}\pi} \int d\nu d\Omega d\bmi{k}\;
    (\mp 2i\nu-1)\cosh\pi\nu \nonumber \\
  & & \qquad\qquad
    {}\times K_{i\nu\pm\frac{1}{2}}(M_k\xi)\;
     e^{-i\bmi{k}\bmi{x}+i\Omega\eta}\,
     \hat{\Psi}_\pm(\nu,\bmi{k},\Omega).
\end{eqnarray}

If we specialize to the massless case,
then after a straightforward calculation,
one finds that
the effective equation of motion for the detector is
\begin{equation}
\left(\frac{\delta_L S_0}{\delta\bar{\hat{\Theta}}}\right)
+{\cal K}_{1/2,n}(\omega)\,\tilde{\hat{\Theta}}(\omega)=0,
\end{equation}
where
\begin{eqnarray}
{\cal K}_{1/2,n}(\omega)&=&
   \frac{a^{n-2}\Gamma(\frac{n-2}{2})}
   {8\pi^{\frac{n}{2}+1}\Gamma(n-2)}
   \int d\nu
   \left[\frac{1}{-\omega/a+\nu+i\epsilon}-
             \frac{1}{\omega/a+\nu-i}\right]
 \nonumber \\
 & & \qquad{}\times
   \frac{(2\nu-i)\cosh\pi\nu}{2\omega/a-i}\,
   \left(\frac{n-3}{2}-i\frac{\omega}{a}\right) 
   (-\gamma^0+\gamma^1) 
 \nonumber \\
 & & \qquad{}\times
   \Gamma\left(\frac{n+1}{2}-1+i\nu\right)
   \Gamma\left(\frac{n-1}{2}-1-i\nu\right)
 \nonumber \\
 & & {}-
   \left[\frac{1}{-\omega/a+\nu+i\epsilon}-
             \frac{1}{\omega/a+\nu+i}\right]
 \nonumber \\
 & & \qquad{}\times
   \frac{(2\nu+i)\cosh\pi\nu}{2\omega/a+i}\,
   \left(\frac{n-3}{2}+i\frac{\omega}{a}\right)
   (\gamma^0+\gamma^1)
 \nonumber \\
 & & \qquad{}\times
   \Gamma\left(\frac{n-1}{2}-1+i\nu\right)
   \Gamma\left(\frac{n+1}{2}-1-i\nu\right).
\end{eqnarray}
Note that the singularity on the real axis of $\nu$
is only at $\nu=\omega/a$.
To avoid this singularity, we add the $+i\epsilon$ term.
(The sign of the $i\epsilon$ term can be determined
by the causality, as above.
Then, the sign is $+$ for 
both $\omega>0$ and $\omega<0$.)
From this $+i\epsilon$ term, 
we can obtain the dissipative coefficient
\begin{eqnarray}
 {\cal R}_{1/2,n}(\omega)&\equiv&
     -\frac{{\rm Im}{\cal K}_{1/2,n}(\omega)}{-\omega} \nonumber \\
 &=& \frac{2^{1-n}\pi^{\frac{3-n}{2}}}{\Gamma(\frac{n-1}{2})}
        \frac{a^{n-1}}{\omega^2}\coth(\pi\omega/a)
     \left|\frac{\Gamma(\frac{n+1}{2}-1+i\frac{\omega}{a})}
         {\Gamma(i\frac{\omega}{a})}\right|^2\; \gamma^0
           \nonumber \\
 &=& \alpha_n\,{\cal R}_{n+1}(\omega)\,
            \coth(\pi\omega/a)\;\gamma^0,
\label{RDR}
\end{eqnarray}
where ${\cal R}_{n+1}(\omega)$ is that of the scalar case
Eq.(\ref{RSR}) in $n+1$ dimensions, and
\begin{equation}
   \alpha_n=\frac{2\sqrt{\pi}\Gamma(\frac{n}{2})}
              {\Gamma(\frac{n-1}{2})}.
\end{equation}
Here, we have again included the extra $-$ sign as a convention
in the definition of Eq.(\ref{RDR}).
We also define 
\begin{equation}
 \bar{{\cal R}}_{1/2,n}(-\omega)\equiv
     -{\cal R}_{1/2,n}(-\omega)=
    \alpha_n{\cal R}_{n+1}(-\omega)\,
            \coth(\pi\omega/a)\;\gamma^0.
\end{equation}

Then, we can use the fluctuation-dissipation theorem
of a fermionic operator~\cite{FujTer98}.
The power spectrum of the Dirac field is defined by
\begin{equation}
  {\cal F}_{1/2,n}(\omega)\equiv\Delta_n^{-1} {\rm Tr}\left[
    \gamma^0\int dt'\;e^{-i\omega(t-t')}
  \langle0_M|\hat{\psi}(\tau)\bar{\hat{\psi}}(\tau')|0_M\rangle
   \right],
\end{equation}
and then,
\begin{equation}
{\rm Tr}\left[\gamma^0
 \langle0_M|\hat{\psi}(\tau)\bar{\hat{\psi}}(\tau)|0_M\rangle\right]
=\frac{\Delta_n}{2\pi}\int_{-\infty}^\infty d\omega\;
 {\cal F}_{1/2,n}(\omega).
\end{equation}
Thus, by using Eq.(\ref{thermal}), 
the theorem~\cite{FujTer98} says that
\begin{eqnarray}
 \int^\infty_{-\infty} d\omega\;{\cal F}_{1/2,n}(\omega) 
 &=& \frac{4}{\Delta_n} {\rm Tr}\Biggl[\gamma^0
    \int^\infty_0 d\omega\;
   \frac{\hbar\omega}{2}\Bigl\{
   \left[1-\frac{1}{e^{\beta\hbar\omega}+1}\right]
   \,{\cal R}_{1/2,n}(\omega)
    \nonumber \\ & & \qquad\qquad\qquad\qquad  
   {}+\left[\frac{1}{e^{\beta\hbar\omega}+1}\right]\,
    \bar{{\cal R}}_{1/2,n}(-\omega) \Bigr\} \Biggr]\nonumber \\
 &=& \frac{2}{\Delta_n} \int^\infty_{-\infty} d\omega\;
   \frac{\hbar\omega}{e^{\beta\hbar\omega}+1}
    {\rm Tr}[\gamma^0\bar{{\cal R}}_{1/2,n}(-\omega)] \nonumber \\
 &=& 2\alpha_n\int^\infty_{-\infty} d\omega\;
   \frac{\hbar\omega{\cal R}_{n+1}(-\omega)}{e^{\beta\hbar\omega}-1}.
\end{eqnarray}
That is, the power spectrum of the Rindler noise of massless
Dirac field in $n$ dimensions is proportional to
that of massless scalar field Eq.(\ref{RSF}) in $n+1$ dimensions.
From the analysis of the scalar case,
we can thus show the statistical inversion.
This result nicely agrees with that of Takagi~\cite{Takagi86}.

%% file: conclude.tex
We have presented a simple and systematic method
to evaluate the Rindler noise.
We have first calculated the dissipative coefficient
{\em explicitly} from the equations of motion
of the detector and the field.
Then, by using the fluctuation-dissipative theorem,
we have obtained the Rindler noise
which is relevant to the analysis of the Unruh effect.
This method is generalized to the Dirac field, by using 
the fermionic fluctuation-dissipation theorem~\cite{FujTer98}.
These results are perfectly consistent 
with previously known results
including apparent statistical inversion in odd dimensions.

Although the Rindler noise can be calculated
{\em directly} from the action of the field
as Takagi~\cite{Takagi86} did,
we emphasize that there are several advantages
in calculating the noise {\em indirectly}.
That is, to introduce the detector,
calculate the dissipative coefficient
and then use the fluctuation-dissipation theorem,
as we have done in this paper.

First of all, it contains only the {\em classical} calculations
expect for the fluctuation-dissipation theorem.
To calculate the dissipative coefficient,
we only have to eliminate the degree of freedom of the field
from classical equations of motion of the detector and field.
Thus, the calculations become much simpler.

Next, this method gives a hint as to
the origin of the apparent statistics inversion effect.
While the dissipative coefficient for the thermal noise
does not depend on the temperature in an ideal setting,
those for the Rindler noise inevitably depends 
on the ``temperature'' (the acceleration).
This temperature-dependence of the dissipative coefficient
destroys the simple Bose (or Fermi) distribution provided by
the fluctuation-dissipation theorem and 
causes the phenomenon of the ``apparent'' inversion
of statistics in odd dimensions~\cite{Takagi86}.
Why does this difference happen?
From our calculation, the answer is obvious.
Note that all the effect of the field on the detector
is represented by the influence functional~\cite{FeyVer63}
which is constructed from the action of the field,
the detector-field interaction and the initial condition
of the field.
(The influence functional for a scalar field
in the Minkowski vacuum, coupled to a uniformly
accelerating DeWitt detector was derived by Anglin~\cite{Anglin93}.)
In the thermal case, the temperature is the initial condition.
On the other hand, in the Rindler case, 
the temperature is in the detector-field interaction
as the acceleration.
Because it is only the action of the field and
the detector-field interaction that are
needed to obtain the effective equation of motion and
thus the dissipative coefficient,
the dissipative coefficient for the thermal noise
does not depend on the temperature in our idealized treatment
but those for the Rindler noise inevitably depends on the temperature.
(Of course, this point of view would not
explain everything about the apparent
statistics inversion effect.)

Finally, we can see that the fermionic
fluctuation-dissipation theorem~\cite{FujTer98}
works properly.
In the context of condensed matter physics,
it is generally difficult to use the fermionic
fluctuation-dissipation theorem because we usually measure
bosonic quantities such as voltage or electric current.
However, we have applied this fermionic version of the
theorem to the Unruh effect
and have obtained the right result 
including apparent statistical inversion in odd dimensions.
Thus, we have confirmed that the fermionic
fluctuation-dissipation theorem is indeed working properly.

Since the relation between Rindler and Minkowski coordinates
is very similar to the relation between
Schwarzschild and Kruskal coordinates,
there is a hope that the above method could
be applicable to the Hawking radiation~\cite{Hawkin75,Wald75} also.
This would be a future work.

%% file: unruh-fd.bbl
\begin{thebibliography}{10}

\bibitem{Unruh76}
W.~G. Unruh, Phys. Rev. {\bf D14},  870  (1976).

\bibitem{Fullin73}
S.~A. Fulling, Phys. Rev. {\bf D7},  2850  (1973).

\bibitem{Davies75}
P.~C.~W. Davies, J. Phys. {\bf A8},  609  (1975).

\bibitem{UnrWal84}
W.~G. Unruh and R.~M. Wald, Phys. Rev. {\bf D29},  1047  (1984).

\bibitem{DeWitt79}
B.~S. DeWitt,  in {\em General Relativity, an {E}instein Centenary Survey},
  edited by S.~W. Hawking and W. Israel (Cambridge University Press, Cambridge,
  1979).

\bibitem{Takagi86}
S. Takagi, Prog. Theor. Phys. Suppl. {\bf 88},  1  (1986),
Many of earlier references on the Unruh effect
are found in this review.

\bibitem{CalWel51}
H.~B. Callen and T.~A. Welton, Phys. Rev. {\bf 83},  34  (1951).

\bibitem{Kubo57}
R. Kubo, J. Phys. Soc. Jpn {\bf 12},  570  (1957).

\bibitem{FeyVer63}
R.~P. Feynman and F.~L. Vernon, Ann. Phys. {\bf 24},  118  (1963).

\bibitem{Ooguri86}
H. Ooguri, Phys. Rev. {\bf D33},  3573  (1986).

\bibitem{FujTer98}
K. Fujikawa and H. Terashima, Phys. Rev. {\bf E58},  7063  (1998),
hep-th/9809048.
See also, K. Fujikawa, Phys. Rev. {\bf E57},  5023  (1998),
hep-th/9802025.

\bibitem{Rindle66}
W. Rindler, Am. J. Phys. {\bf 34},  1174  (1966).

\bibitem{GraRyz80}
I.~S. Gradshteyn and I.~M. Ryzhik, {\em Table of Integrals, Series,
  and Products} (Academic Press,1980), p. 693, Eq. 6.576--4.

\bibitem{Israel76}
W. Israel, Phys. Lett. {\bf 57A},  107  (1976).

\bibitem{TroVan79}
W. Troost and H. Van~Dam, Nucl. Phys. {\bf B152},  442  (1979).

\bibitem{UnrWei84}
W.~G. Unruh and N. Weiss, Phys. Rev. {\bf D29},  1656  (1984).

\bibitem{HoHoYa85}
M. Horibe, A. Hosoya, and N. Yamamoto, Prog. Theor. Phys. {\bf 74},  1299
  (1985).

\bibitem{ChrDuf78}
S.~M. Christensen and M.~J. Duff, Nucl. Phys. {\bf B146},  11  (1978).

\bibitem{Sewell82}
G.~L. Sewell, Ann. Phys. {\bf 141},  201  (1982).

\bibitem{MarSch59}
P.~C. Martin and J. Schwinger, Phys. Rev. {\bf 115},  1342  (1959).

\bibitem{SoMuGr80}
M. Soffel, B. M{\"u}ller, and W. Greiner, Phys. Rev. {\bf D22},  1935  (1980).

\bibitem{Anglin93}
J.~R. Anglin, Phys. Rev. {\bf D47},  4525  (1993),
hep-th/9210035.

\bibitem{Hawkin75}
S.~W. Hawking, Commun. Math. Phys. {\bf 43},  199  (1975).

\bibitem{Wald75}
R.~M. Wald, Commun. Math. Phys. {\bf 45},  9  (1975).

\end{thebibliography}
